\documentclass[graybox]{svmult}

\usepackage{mathptmx}       % selects Times Roman as basic font
\usepackage{helvet}         % selects Helvetica as sans-serif font
\usepackage{courier}        % selects Courier as typewriter font
\usepackage{type1cm}        % activate if the above 3 fonts are
\usepackage{makeidx}         % allows index generation
\usepackage{graphicx}        % standard LaTeX graphics tool
\usepackage{multicol}        % used for the two-column index
\usepackage[bottom]{footmisc}% places footnotes at page bottom

\makeindex             % used for the subject index
\begin{document}

\title*{the Void Galaxy Survey}
\author{R. van de Weygaert, K. Kreckel, E. Platen, B. Beygu, J. H. van Gorkom, 
J. M. van der Hulst,  M. A. Arag\'on-Calvo, P. J. E. Peebles, T. Jarrett, G. Rhee, 
K. Kova\v{c}, C.-W. Yip}
\authorrunning{Van de Weygaert et al.} 
\institute{R. van de Weygaert \at Kapteyn Astronomical Institute, Univ. Groningen, P.O. Box 800, 9700AV Groningen, the Netherlands \email{weygaert@astro.rug.nl}}
%\institute{R. van de Weygaert, E. Platen, B. Beygu, J.M. van der Hulst \at Kapteyn Astronomical Institute, Univ. Groningen, Groningen, the Netherlands
%\and K. Kreckel, J. van Gorkom \at Dept. of Astronomy, Columbia University, New York, NY 10027, USA
%\and M. Arag\'on-Calvo, C.-W. Yip \at The Johns Hopkins University, 3701 San Martin Drive, Baltimore, MD 21218, USA
%\and P.J.E. Peebles \at Joseph Henry Laboratories, Princeton University, Princeton, NJ 08544, USA
%\and T. Jarret \at IPAC, Spitzer Science Center, JPL, Caltech, Pasadena, CA 91125, USA 
%\and G. Rhee \at Dept. Physics \& Astronomy, UNLV, Las Vegas, NV 89154-4002, USA
%\and K. Kova\v{c} \at Max Max-Planck-Institut f\"ur Astrophysik, D-85748 Garching, Germany}
%\institute{Name of First Author \at Name, Address of Institute, \email{name@email.address}
%\and Name of Second Author \at Name, Address of Institute \email{name@email.address}}
\maketitle

% Too much empty space in the original style file!
\vskip-1.2truein

\abstract{The Void Galaxy Survey (VGS) is a multi-wavelength program to study $\sim$60 void galaxies. 
Each has been selected from the deepest interior regions of identified voids 
in the SDSS redshift survey on the basis of a unique geometric technique, with no a prior 
selection of intrinsic properties of the void galaxies. 
The project intends to study in detail the gas content, star formation history and 
stellar content, as well as kinematics and dynamics of void galaxies and their companions in 
a broad sample of void environments. It involves the HI imaging of the gas distribution 
in each of the VGS galaxies. Amongst its most tantalizing findings is the possible evidence 
for cold gas accretion in some of the most interesting objects, amongst which are a 
polar ring galaxy and a filamentary configuration of void galaxies. Here we shortly describe the 
scope of the VGS and the results of the full analysis of the pilot sample of 15 void galaxies.}
\section{Introduction: Voids and Void Galaxies}
\label{sec:1}
Voids have been known as a feature of the Megaparsec universe since the first galaxy redshift surveys were compiled 
\cite{einasto1980,kirshner1981,lapparent1986}.  {\it Voids} are enormous regions with sizes in the range of $20-50h^{-1}$ Mpc that are practically devoid 
of any galaxy, usually roundish in shape and occupying the major share of space in the Universe \cite{weyplaten2009}. Forming an essential ingredient 
of the {\it Cosmic Web} \cite{bondweb1996}, they are surrounded by elongated filaments, sheetlike walls and dense compact clusters.  

A major point of interest is that of the galaxies populating the voids. The pristine environment of 
voids represents an ideal and pure setting for the study of galaxy formation. Largely unaffected by the complexities and 
processes modifying galaxies in high-density environments, the isolated void regions must hold important clues to the formation 
and evolution of galaxies. This makes the relation between {\it void galaxies} and their surroundings an important aspect 
of the interest in environmental influences on galaxy formation  \cite{szomoru1996,grogin1999,hoyvog2002,patiri2006,peebnuss2010}. 

\begin{figure}[b]
\centering
%\vskip -0.5truecm
\mbox{\hskip -1.0truecm\includegraphics[width=1.135\textwidth]{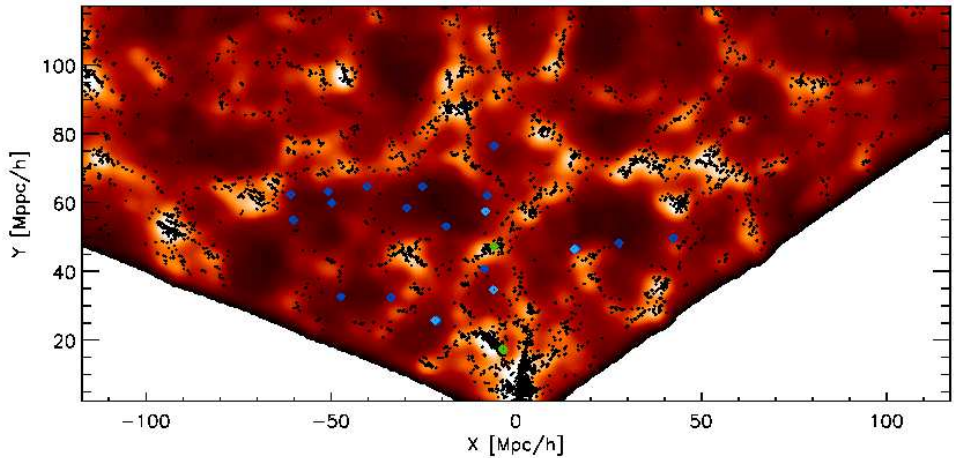}}
\caption{SDSS density map and identification of voids in the SDSS galaxy  redshift survey region from which we selected 
the galaxies in the Void Galaxy Survey, in a slice of thickness 4$h^{-1}$Mpc. The DTFE computed galaxy density
  map, Gaussian smoothed on a scale of $R_f=1h^{-1}$Mpc, is represented by
  the colorscale map. The SDSS galaxies are superimposed as dark dots. Blue diamonds: VGS pilot sample galaxies. 
  Dark blue diamonds: VGS void galaxies from the full sample. Green diamonds: control sample galaxies. From Kreckel et al. 2011.
\label{fig:sdssmapvoid}}
\vskip -0.5truecm
\end{figure}
Amongst the issues relevant for our understanding of galaxy and structure formation, void galaxies have posed several 
interesting riddles and questions. Of cosmological importance is the finding from optical and HI surveys that 
the density of faint galaxies in voids is only $1/100$th that of the mean. As has been strongly emphasized by 
Peebles \cite{peebles2001}, this dearth of dwarf void galaxies cannot be straightforwardly understood in our standard 
$\Lambda$CDM based view of galaxy formation: voids are expected to be teeming with dwarfs and low surface brightness 
galaxies. Various astrophysical processes, ranging from gas and radiation feedback processes to environmental properties of 
dark matter halos, have been suggested \cite{peebles2001,mathis2002,hoeft2006,furlanetto2006,tinker2009}. The issue 
is, however, far from solved and progress will depend largely on new observations that characterize void galaxies and their immediate 
environment. An additional issue of cosmological interest is whether we can observe the intricate filigree of substructure in voids, 
expected as the remaining debris of the merging voids and filaments in the hierarchical formation process \cite{dubinski1993,
weygaert1993,gottloeber2003,shethwey2004}. 

Of particular interest in the present context is the manifest environmental influence on the nature of void galaxies. They 
are found to reside in a more youthful state of star formation. As a population, void galaxies are statistically bluer, have a later 
morphological type, and have higher specific star formation rates than galaxies in average density environments \cite{grogin1999,
rojas2004,patiri2006}. Whether void galaxies are intrinsically different or whether their characteristics 
are simply due to the low mass bias of the galaxy luminosity function in low density regions is still an issue of discussion. 

An important aspect towards understanding the nature of void galaxies is that of their gas content, about which far less is known 
than their stellar content. The early survey of 24 IRAS selected IRAS galaxies within the Bo\"otes void by \cite{szomoru1996} 
revealed that most of them were gas rich and disk-like, with many gas rich companions. Fresh gas accretion is necessary for galaxies to 
maintain star formation rates seen today without depleting their observed gas mass in less than a Hubble time \cite{larson1972}. 
Historically, this gas was assumed to condense out of reservoirs of hot gas existing in halos around galaxies \cite{rees1977,white1991}, 
with some amount of gas recycling  via galactic fountains \cite{fraternali2008}.  However, recent 
simulations have renewed interest in the slow accretion of cold gas along filaments \cite{binney1977,keres2005,dekel2006}. 

The unique nature of void galaxies provides an ideal chance to distinguish the role of environment in gas accretion and 
galaxy evolution on an individual basis. Amongst others, their inherent isolation may allow us to distinguish the effects 
of close encounters and galaxy mergers from other mechanisms of gas accretion.

\section{the Void Galaxy Survey}
\label{sec:2}
The Void Galaxy Survey (VGS) is a multi-wavelength study of $\sim$60 void galaxies, geometrically selected from the SDSS 
galaxy redshift DR7 survey database.  The project has the intention to study in detail the gas content, star formation history and 
stellar content, as well as kinematics and dynamics of void galaxies and their companions in a broad sample of void environments.  
Each of the 60 galaxies has obtained a VGS number, VGS1 until VGS60. Ultimately, we aim to compile a sample of 50-100 void galaxies.  

\begin{figure}[h]
  \centering
  \vskip -0.25truecm
  \includegraphics[width=1.0\linewidth]{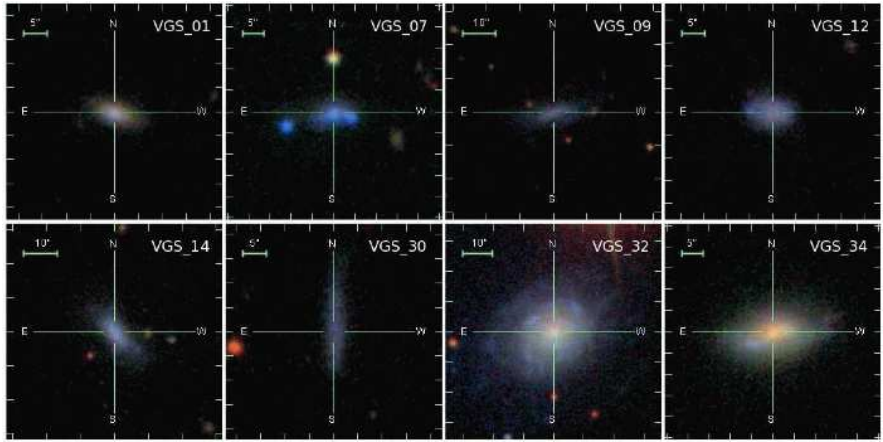}
  \caption{A selection of 8 VGS void galaxies from the SDSS DR7 galaxy redshift survey. These 
galaxies are part of the VGS pilot survey. The images, composite color images from the SDSS Finding Chart tool, 
are scaled to the same scale. From Kreckel et al. 2011.}
\vskip -0.25truecm
  \label{fig:sdsssam}
\end{figure}

All galaxies have been selected from the deepest interior regions of identified voids in the SDSS redshift survey on the 
basis of a unique  geometric technique, with no a priori selection on intrinsic magnitude, color or morphology of the void galaxies. 
The most isolated and emptiest regions in the Local Universe are obtained from the galaxy density and structure maps produced by 
the DTFE/spine reconstruction technique \cite{schaapwey2000,weyschaap2009,platen2007,aragon2010}. From the spatial distribution 
of the SDSS galaxies, in a volume from $z=0.003$ to $z=0.03$, we reconstruct a density field by means of the DTFE procedure, the Delaunay 
Tessellation Field Estimator . In addition to the computational efficiency of the procedure, 
the density maps produced by DTFE have the virtue of retaining the anisotropic and hierarchical structures which are so characteristic 
of the Cosmic Web. The Watershed Void Finder is applied to the DTFE density field for identifying its underdense void basins.

\begin{figure}[h]
\centering
\vskip -0.25truecm
\mbox{\hskip -0.95truecm\includegraphics[width=1.1\linewidth]{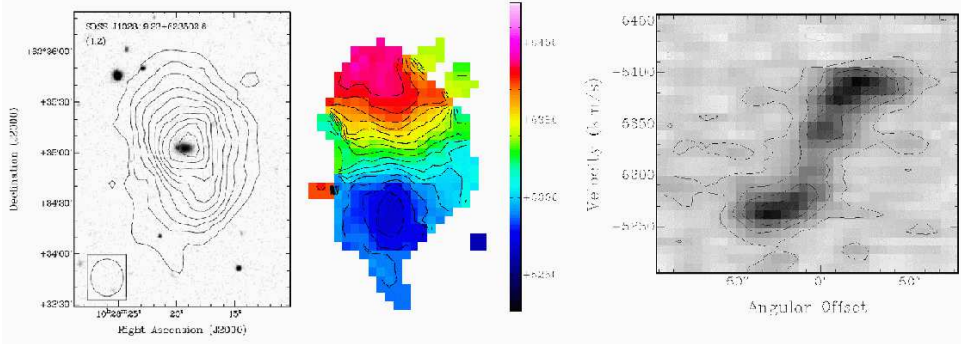}}
\caption[]{Targeted void galaxy VGS12, the polar ring void galaxy. Left: HI intensity map, superimposed on optical image. 
Contours are at $5 \times 10^{19} cm^{-2}$, plus increments of $10^{20} cm^{-2}$. Centre: velocity field. Lines indicate 
increments of 8 km s$^{-1}$. Right: position-velocity diagram along the kinematic major axis. From Kreckel et al. 2011.
\label{fig:polarring}}
\vskip -0.5truecm
\end{figure}

Using the WSRT we have thus far mapped the HI structure of 55 of the 60 galaxies. Of the total VGS sample of 60 
void galaxies, the pilot subsample of 15 galaxies has been fully analyzed \cite{stanonik2009,kreckel2011}. 
A necessary sample of comparison galaxies is obtained through simultaneous coverage of regions in front and behind the targeted 
void galaxies, probing the higher density regions surrounding the targeted void. Note that existing blind HI surveys (alfalfa, HIPASS) are 
limited in not resolving the tell-tale HI structures found in the VGS. 

In addition to the 5-band photometry and spectroscopy from the SDSS, we obtain deep B- and R-band imaging of all sample galaxies 
with the La Palma INT telescope and high resolution slit spectroscopy of a subsample of 
our VGS galaxies. The deep imaging allows us to detect low surface brightness features such as extended, unevolved, stellar disks, 
tidal streams, the stellar counterparts of several detected HI features (polar rings, tails, etc.) and of the faint HI dwarf 
companions. Such information is crucial for distinguishing intrinsic formation and evolution scenarios from external processes such as 
merging and tidal interactions. To probe the old stellar population of the VGS void galaxies, for 10 galaxies we have obtained near-IR JHK 
WIRC imaging at the 5-meter Palomar Hale telescope. In order to assess the distribution of star formation and associated star formation rates, 
we are obtaining GALEX UV data of 45 galaxies, and have obtained H$\alpha$ imaging of the complete sample at the MDM telescope. 

\section{Results of the VGS survey: current state of affairs} 
The first results of the Void Galaxy Survey are tantalizing and has revealed a few surprising gas configurations. 
With a HI mass limit of $\sim 2 \times 10^7$ M$_{\odot}$ and column density limit of $\sim 10^{19}$cm$^{-2}$, 
our HI survey provides a significantly improved view of HI in void galaxies compared to past studies \cite{szomoru1996}. 
Figure~\ref{fig:polarring} shows one of the most surprising specimen in our survey, the polar ring void galaxy 
VGS12 (see sect.~\ref{sec:elite}). It is one of the relatively large number of void galaxies, possibly together with e.g.  
KK246 and VGS31, that show evidence for cold mode accretion.

The first fully analyzed sample of 15 void galaxies demonstrated the success of our strategy \cite{kreckel2011}. 
With 14 detections out of 15, the HI detection rate is very high. We discovered one previously known and five previously unknown 
companions, while two appear to be interacting. Of these five befriended void galaxies, two are interacting in HI.  All HI-detected 
companions have optical counterparts within the SDSS.  Of the nine isolated void galaxies, many exhibit irregularities in the kinematics 
of their gas disks. Based on their ~150 km s$^{-1}$ velocity width, the detected target galaxies have a range of HI masses from 
$0.35-3.8\times10^9$ M$_{\odot}$,  Companion galaxies have masses ranging from $0.5-4.5\times 10^8$ M$_{\odot}$.

While our targeted void galaxies are small, they would not be classified as dwarf galaxies \cite{hodge1971}.  All have 
M$_r$ $<$ -16 and exhibit small circular velocities of 50-100 km/s. All exhibit signs of rotation, though limiting resolution and lower sensitivity 
at the disk outskirts means we do not always see a flattening of the rotation curve.  M$_{dyn}$ is typically 10$^9$-10$^{10} M_{\odot}$. 
The detected companions are more dwarfish.

The void galaxy population appears to represent the extreme blue and faint tail of an otherwise normal galaxy population. 
There are a few characteristics which seem to set them apart, mainly concerning their HI gas content and star formation 
activity. 

\begin{figure}[h]
\centering
\vskip-0.0cm
\mbox{\hskip -1.2truecm\includegraphics[width=1.15\linewidth]{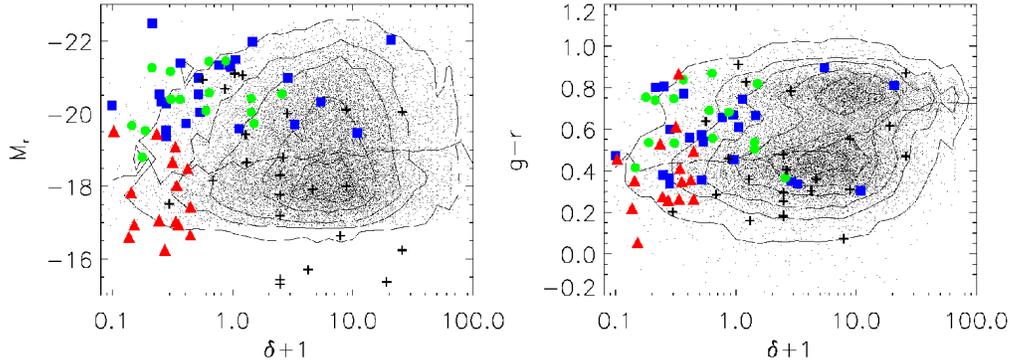}}
\vskip -0.25truecm
\caption{Magnitudes and colour of void galaxies as a function of density excess/deficit $\delta$. 
Left: distribution of r-b absolute magnitudes for our void galaxy sample (triangles),  the Bo\"otes 
void galaxy sample of Szomoru et al. (1996) (blue squares) and the CfA void galaxy sample of Grogin \& 
Geller (1999) (crosses). These are compared to the general colour-magnitude diagram of a volume-limited 
sample of SDSS galaxies, with $z<0.02$ and $M_r<-16.9$ (dots). From Kreckel et al. 2011. 
\label{fig:colmag}}
\vskip -0.5cm
\end{figure}
\subsection{Magnitudes and Colors}
Despite their affected outer regions, our study finds that in the colour-magnitude diagram the target void galaxies 
nicely reside at the faint end of the blue cloud of galaxies. This can be immediately appreciated from the diagrams 
in figure~\ref{fig:colmag}, showing the magnitude and color distribution of our galaxies as a function of density 
excess/deficit $\delta$. Most of our galaxies find themselves on the blue sequence of SDSS galaxies, towards the bluest 
edge of these galaxies. We find that our pilot sample galaxies are at the 
faint end of the galaxy luminosity function (see e.g. also \cite{rojas2004}) !  Because our geometric selection procedure 
manages to probe the extremely underdense and 
desolate void interiors, our void galaxy sample is able to probe specifically those low luminosity galaxies which make up 
the bulk of the void galaxy population and were previously inaccessible (fig~\ref{fig:colmag}, left).  Also apparent is 
the dominance of blue galaxies at the deepest underdensities (fig.~\ref{fig:colmag}, right).

\subsection{Star Formation and Gas Content}
A key aspect of the HI observations is that even in these rather desolate and underdense void regions 
it revealed several cases of very irregular HI morphologies, marked by {\it features} such as disturbed 
HI disks, tails, warps, and cold gas filaments. This suggests that void galaxies are activily building up. 
A related second key aspect is the high abundance of faint {\it companions}, non-interacting as well as merging. 

\begin{figure}[h]
\centering
\vskip -0.0truecm
\mbox{\hskip -0.6truecm\includegraphics[width=1.08\linewidth]{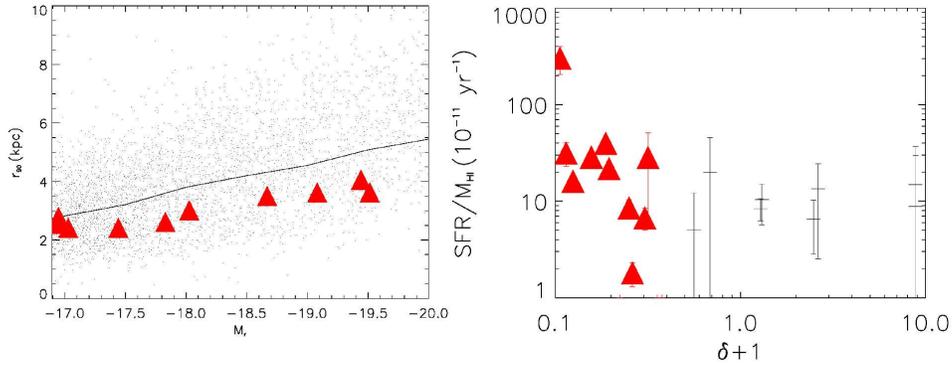}}
\vskip -0.25truecm
\caption{Two deviant characteristics of VGS void galaxies. Left: The $r$-band r$_{90}$ radii of the stellar disk.  
Our late-type void galaxies (triangles) fall systematically below the median (line) of a volume limited SDSS sample of 
late-type galaxies (dots). Right: star formation rate per hydrogen mass, SFR/$M_{HI}$, as a function of density of our void 
galaxies (triangles) and our control sample (crosses).  The void galaxies have a higher star formation rate per hydrogen 
mass. From Kreckel et al. 2011.
\label{fig:sfrhi}}
\vskip -0.25truecm
\end{figure}

One particular aspect in which we find a systematic deviation of our void galaxies, with respect to the norm for 
similar galaxies, is their size. They have stellar disks that are smaller than average, 
with the $r$-band r$_{90}$ radius of our late type void galaxies systematically lower than the median for late 
type galaxies (fig.~\ref{fig:sfrhi}, left). However, the result is tentative and might be beset by a hidden 
selection effect. 

Perhaps the most outstanding characteristic of void galaxies is that of their star formation properties. In general, the 
stellar and star formation properties of our VGS pilot sample are in agreement with the values found in other samples 
of void galaxies \cite{rojas2004}. In this respect, it is relevant that the HI mass of the VGS galaxies appears to be 
typical in following the global trend of an increasing hydrogen mass $M_{HI}$ as their optical (r-band) luminosity 
decreases: the smallest galaxies have been less efficient at turning gas into stars. However, when assessing possible 
trends with density, we find that the specific star formation rate (SFR per stellar mass) of the galaxies displays 
a suggestive systematic trend. There is a distinct trend for an increase of the star formation rate per HI mass 
for galaxies in lower density areas (fig.~\ref{fig:sfrhi}, right). 

In all, we find that the outer regions and immediate environment of void galaxies testifies of strong recent 
interactions and star formation activity. This, by itself, is a surprising finding for galaxies populating the 
most desolate areas of our Universe. 

\begin{figure}[h]
\centering
\mbox{\hskip -0.95truecm\includegraphics[width=1.1\linewidth]{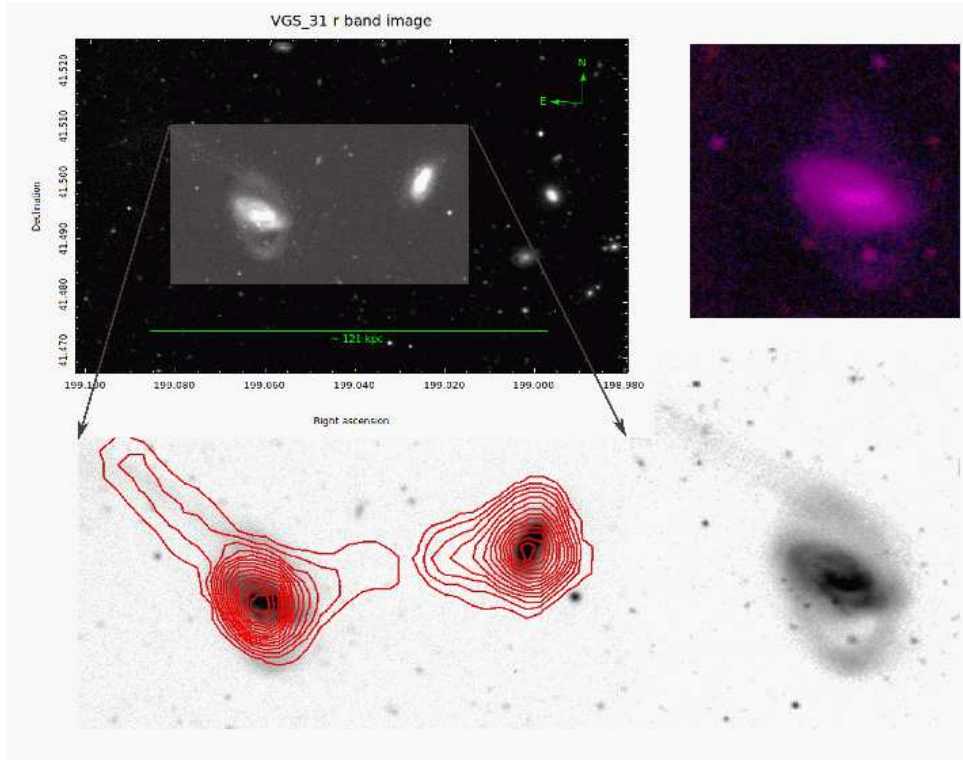}}
\caption[]{The elongated void galaxy complex VGS31. Top left: INT r-band image of the complex. It consists of Markarian galaxy VGS31b (left), 
the VGS target galaxy VGS31a (centre) and the faint galaxy VGS31c (right). Zooming in on the central region with VGS31a and VGS31b, the 
bottom panel displays the HI intensity contour map superimposed on the r-band image. The intricate tails and/or streams around VGS31a, 
as well as the rather distorted interior of VGS31b are clearly visible in the r-band image zoom-in  (lower rigthand corner). The old 
stellar population surfaces in th J+K image in the top righthand panel. From Beygu et al. 2011.
\label{fig:vgs31}}
\vskip -0.0truecm
\end{figure}

\subsection{Exuberance in the Desert}
\label{sec:elite}
An outstanding specimen of our sample is the polar disk galaxy VGS12 \cite{stanonik2009}. Amongst the 
most lonely galaxies in the universe, it has a massive, star-poor HI disk that is perpendicular to the disk of the 
central void galaxy. No optical counterpart to the HI disk has yet been found, even though the inner 
optical galaxy is actively forming stars. The galaxy is located within a tenuous wall in between two large roundish voids.
The undisrupted appearance of the original stellar disk renders a merger origin unlikely. It suggests slow accretion of 
cold gas \cite{binney1977,keres2005,dekel2006}, at the crossing point of the outflow from the 
two voids. Cold accretion as a formation mechanism for polar ring galaxies has been seen to occur in simulations 
\cite{maccio2006,brook2008}.

Another fascinating object is VGS31. It defines a system of three galaxies, stretching 
out over 57 kpc and possibly connected by a HI bridge. The easternmost object is a Markarian 
galaxy, marked by prominent stellar streams wrapping around the central galaxy 
and a separate tidal tail or stream (see fig.~\ref{fig:vgs31}). These might be the remnants of 
the recent infall of one or two satellite galaxies \cite{beygu2011}. The westernmost object is considerably 
fainter than the other two galaxies. The tails and streams 
are also visible in the recent deep INT B imaging as well as on the NIR J and K maps, possibly 
with a slight and unique misalignment. The fact that the gas and all objects 
involved appear to be stretched along a preferred direction may be suggestive of a  system 
situated within a tenuous filament within the large encompassing void.

\section{Conclusions}
With the analysis of the first 15 galaxies completed, and the analysis of 45 additional ones in progress, we find that the 
VGS void galaxies have small optical stellar disks and typical HI masses for their luminosity.  Consistent with previous surveys, 
they are bluer and have increased rates of star formation, with the suggestion of a trend towards increased star formation at 
lowest density.   While this pilot sample is too small for any statistical findings, we did discover many of our targets to be 
individually interesting dynamically and kinematically in their HI properties. In particular, a few show direct evidence 
of ongoing cold mode accretion. Ultimately, we aim to compile a sample of 50-100 void galaxies. 

\begin{acknowledgement}
Over the past years many useful and encouraging discussions shaped our project. To this end we wish 
to particularly thank Michael Vogeley, Ravi Sheth, Changbom Park, Bernard Jones, Wojciech Hellwing and 
Reynier Peletier.
\end{acknowledgement}

%%% Bibliography

%\input{referenc}

\begin{thebibliography}{99.}%
% Use the following syntax and markup for your references if 
% the subject of your book is from the field 
% "Mathematics, Physics, Statistics, Computer Science"
%
\bibitem{aragon2010} Arag{\'o}n-Calvo M.A., Platen E., van de Weygaert R., Szalay A.S., 2010, ApJ, 723, 364
\bibitem{beygu2011} Beygu B., et al. , 2011, MNRAS, in prep.
\bibitem{binney1977} Binney, J. 1977, ApJ, 215, 483
\bibitem{bondweb1996} Bond J.~R., Kofman L., Pogosyan, D. 1996, Nature, 380, 603
\bibitem{brook2008} Brook C.B., et al., 2008, ApJ, 689, 678
\bibitem{dekel2006} Dekel A., Birnboim Y. 2006, MNRAS, 368, 2
\bibitem{dubinski1993} Dubinski J., da Costa L.~N., Goldwirth D.~S., Lecar M., Piran T. 1993, ApJ, 410, 458
\bibitem{lapparent1986} de Lapparent V., Geller M.~J., Huchra J.~P. 1986, ApJL, 302, L1
\bibitem{einasto1980} Einasto J., Joeveer M., Saar E. 1980, MNRAS, 193, 353
\bibitem{fraternali2008} Fraternali F., Binney J.~J. 2008, MNRAS, 386, 935
\bibitem{furlanetto2006} Furlanetto S.~R., Piran T., 2006, MNRAS , 366, 467
\bibitem{gottloeber2003} Gottl{\"o}ber S., {\L}okas E.~L., Klypin A., Hoffman Y., 2003, MNRAS, 344, 715
\bibitem{grogin1999} Grogin N.~A., Geller M.~J. 1999, AJ, 118, 2561
\bibitem{hodge1971} Hodge P.W., 1971, Ann. Rev. Astron. Astrophys., 9, 35
\bibitem{hoeft2006} Hoeft M., Yepes G., Gottl{\"o}ber S., Springel V., 2006, MNRAS, 371, 401
\bibitem{hoyvog2002} Hoyle F., Vogeley M.~S. 2002, ApJ, 566, 641
\bibitem{keres2005} Kere{\v s} D., Katz N., Weinberg D.~H., Dav{\'e} R., 2005, MNRAS, 363, 2
\bibitem{kirshner1981} Kirshner R.~P., Oemler Jr. A., Schechter P.~L., Shectman S.~A., 1981, ApJL, 248, L57
\bibitem{kreckel2011} Kreckel K., Platen E., Arag\'on-Calvo M. A., van Gorkom J. H., van de Weygaert R., 
van der Hulst J.M., Kova{\u c} K., Yip C.-W., Peebles, P.J.E., 2011, AJ, 141, 4
\bibitem{larson1972} Larson R.~B. 1972, Nature, 236, 21
\bibitem{maccio2006} Maccio A.V., Moore B., Stadel J., 2006, ApJ, 636, 25
\bibitem{mathis2002} Mathis H., White S.~D.~M., 2002, MNRAS, 337, 1193
\bibitem{park2007} Park C., Choi Y.-Y., Vogeley M.~S., Gott J.~R.~I., Blanton M.~R. 2007, ApJ, 658, 898
\bibitem{patiri2006} Patiri S.~G., Betancort-Rijo J.~E., Prada F., Klypin A., Gottl{\"o}ber S., 2006, MNRAS, 369, 335
\bibitem{peebles2001} Peebles P.~J.~E. 2001, ApJ, 557, 495
\bibitem{peebnuss2010} Peebles P.~J.~E., Nusser A., 2010, Nature, 465, 565
\bibitem{platen2007} Platen E., van de Weygaert R., Jones B.~J.~T. 2007, MNRAS, 380, 551
\bibitem{rees1977} Rees M.~J., Ostriker J.~P., 1977, MNRAS, 179, 541
\bibitem{rojas2004} Rojas R.~R., Vogeley M.~S., Hoyle F., Brinkmann J., 2004, ApJ, 617, 50
\bibitem{schaapwey2000} Schaap W.~E., van de Weygaert R., 2000, A\&A, 363, L29
\bibitem{shethwey2004} Sheth R.~K., van de Weygaert R., 2004, MNRAS, 350, 517
\bibitem{stanonik2009} Stanonik K., Platen E., Arag\'on-Calvo M. A., van Gorkom J. H., van de Weygaert R., 
van der Hulst J. M., Peebles P. J. E., 2009, ApJL, 696, L6
\bibitem{szomoru1996} Szomoru A., van Gorkom J.~H., Gregg M.~D., Strauss M.~A., 1996, AJ, 111, 2150
\bibitem{tinker2009} Tinker J.~L., Conroy C. 2009, ApJ, 691, 633
\bibitem{weygaert1993} van de Weygaert R., van Kampen E., 1993, MNRAS, 263, 481
\bibitem{weyschaap2009} van de Weygaert R., Schaap E., 2009, in LNP 665, Data Analysis in Cosmology. eds. V. J. Mart\'{\i}nez, E. Saar, 
E. Mart\'{\i}nez-González, M.-J. Pons-Border\'{\i}a. Springer, p.291-413
\bibitem{weyplaten2009} van de Weygaert R., Platen E., 2009, Modern Phys. Lett. A., in press, arXiv:0912.2997
\bibitem{white1991} White S.~D.~M., Frenk C.~S., 1991, ApJ, 379, 52
\end{thebibliography}
\end{document}